\documentclass[fleqn, usenatbib, useAMS]{mnras}

\usepackage{graphicx}   
\usepackage{color}
\usepackage{lscape}
\usepackage{txfonts}
\usepackage{amsfonts}

\title[$\Gamma$-$\Re$ correlation in AGNs]
{The condensation of the corona for the correlation between the hard X-ray photon index $\Gamma$
and the reflection scaling factor $\Re$ in active galactic nuclei}

\author[Erlin Qiao and B.F. Liu]{Erlin Qiao $^{1,2}$\thanks{E-mail:
qiaoel@nao.cas.cn} and B.F. Liu $^{1,2}$\\
$^{1}$Key Laboratory of Space Astronomy and Technology, National Astronomical Observatories, Chinese Academy of
Sciences, Beijing 100012, China \\
$^{2}$School of Astronomy and
Space Sciences, University of Chinese Academy of Sciences, 19A Yuquan Road, Beijing 100049, China\\}

\date{Accepted XXX. Received YYY; in original form ZZZ}

\pubyear{2016}

\begin{document}
\label{firstpage}
\pagerange{\pageref{firstpage}--\pageref{lastpage}}
\maketitle
\begin{abstract}
Observationally, it is found that there is a strong correlation between the hard X-ray photon index $\Gamma$ and
the Compton reflection scaling factor $\Re$ in active galactic nuclei.
In this paper, we propose that the $\Gamma-\Re$ correlation can be explained  within the framework of the
condensation of the hot corona onto the cold accretion disc around a supermassive black hole.
In the model, it is presumed that, initially, a vertically extended hot gas (corona) is supplied to the central
supermassive black hole by capturing the interstellar medium and stellar wind.
In this scenario, when the initial mass accretion rate $\dot M/ \dot M_{\rm Edd} \gtrsim 0.01$,  
at a critical radius $r_{\rm d}$, part of the hot gas begins to condense onto the
equatorial disc  plane of the black hole, forming an inner cold accretion disc.
Then the  matter is accreted in the form of the disc-corona structure extending down to the 
innermost stable circular orbits of the black hole. 
The size of the inner disc is determined by the initial mass accretion rate.
With the  increase of the initial mass accretion rate, the size of the inner disc increases, which results in
both the increase of the Compton reflection scaling factor $\Re$ and the increase of the hard X-ray photon index $\Gamma$.
By comparing with a sample of Seyfert galaxies with well-fitted X-ray spectra,
it is found that our model can roughly explain the observations. Finally, we discuss the possibility to apply our  
model to high mass X-ray binaries, which are believed to be fueled by the hot wind from the companion star.
\end{abstract}




\section{Introduction}
The hard X-ray spectra of bright Seyfert galaxies can phenomenologically be described by a power law with a 
high energy exponential cut off at a few hundred keV, i.e.,  
$F(E)\propto E^{-\Gamma+1}{\rm{exp}}(-E/E_{c})$ (with $\Gamma$ being the hard X-ray
photon index, and $E_{\rm c}$ being the e-folding cut-off energy), plus a reflection component \citep[e.g.][]{Nandra1994}.
It is believed that the hard power-law X-ray spectrum is produced by the inverse Compton scattering 
of the soft photons from a cold accretion disc in an optically thin, hot corona (Sunyaev \& Titarchuk 1980).
While the reflection spectrum is generally thought to be produced by
the illumination of the cold disc by the hot corona
\citep{White1988,George1991,Laor1991,Done1992,Magdziarz1995,Ross2005,Dauser2010,Garcia2011,Garcia2013}.
The relative strength of the reflection component depends on the fraction of the emission of 
the hot corona intercepted by the cold accretion disc, which is often normalized by a reflection scaling 
factor $\Re$, defined as $\Re=\Omega/{2\pi}$ (with $\Omega$ being the solid angle 
covered by the accretion disc as seen from the hot corona). 
Originally, it was found that there is positive correlation between $\Gamma$ and $\Re$ by analyzing the
observational data of Ginga for black hole X-ray transient GX 339-4 in the low/hard state \citep{Ueda1994},
which was confirmed by analyzing the RXTE/PCA data in the low/hard 
state of GX 339-4 during 1996-1997 \citep{Revnivtsev2001}.
A similar correlation between $\Gamma$ and $\Re$ was also found in black hole high mass X-ray binary Cyg X-1 
\citep[][for review]{Gilfanov1999,Ibragimov2005,Gilfanov2010}.
The first report on the  correlation between $\Gamma$ and $\Re$ in active galactic nuclei (AGNs) was 
by \citet{Magdziarz1998}, in the type 1  Seyfert galaxy NGC 5548.
Later, this correlation was confirmed by a comprehensive study of a sample, including Seyfert galaxies, radio 
galaxies and black hole X-ray binaries and weakly magnetized neutron stars
\citep{Zdziarski1999,Zdziarski2003}.
A recent study on a sample of 28 Seyfert galaxies further confirmed this correlation \citep{Lubinski2016}.

The correlation between $\Gamma$ and $\Re$ is suggested to be used to explore the geometry 
of the accretion flow around a supermassive black hole in AGNs \citep[e.g.][]{Zdziarski1999}.
Generally, it is believed that, for luminous AGNs, mainly including the  bright 
Seyfert galaxies and radio quiet quasars, the accretion mode is a cold accretion disc sandwiched  
by a hot corona extending down to the innermost stable circular orbits (ISCO) of the central  black hole
\citep{Shakura1973,Shields1978,Malkan1982,Elvis1994,Kishimoto2005,Shang2005}. 
In such a disc-corona model, the cold accretion disc mainly 
contributes to the optical/UV emission, and the hot corona mainly contributes to the X-ray emission.
Meanwhile, if the hot corona is assumed to be moving away from or towards the cold accretion disc with a mildly
relativistic bulk velocity, the model can well explain the observed
correlation between the hard X-ray photon  index $\Gamma$
and the reflection scaling factor $\Re$ \citep{Beloborodov1999,Malzac2001}.
However, theoretically, in the disc-corona model, the formation 
of the corona is a big problem. Historically, it was only assumed that a fraction of the matter in the disc
is transfered to the coronal region to explain the observed strong X-ray emissions in luminous AGNs
\citep[e.g.][]{Haardt1991,Haardt1993,Svensson1994,Stern1995}.
Although some important progresses have been achieved 
via magnetohydrodynamic (MHD) simulations in the past few decades, it is still in debate currently
\citep[e.g.][]{Miller2000,Hirose2006,Bai2013,Fromang2013,Uzdensky2013,Takahashi2016}.
\citet{Jiang2014} performed a three dimensional 
radiation MHD simulation, they found that the fraction of the  energy dissipated 
in the corona increases with decreasing the surface density of the accretion disc. However, 
with a lowest surface density in their simulation, the maximum fraction of the energy dissipated 
in the corona is only 3.4$\%$, which is inconsistent with
the strong X-ray emission often observed in luminous AGNs
\citep{Vasudevan2007,Vasudevan2009}.

As suggested by \citet{Liu2015}, the physical properties of the initial gas fuel is important for the 
X-ray emission for luminous AGNs. In the paper, we interpret the $\Gamma-\Re$ correlation 
in luminous AGNs within the framework of the 
condensation of the hot corona onto the cold accretion disc around a supermassive black hole.
In this model, it is presumed that, initially, the matter is accreted in the form of vertically extended hot gas (corona)
at the region beyond the Bondi radius of the black hole.  
It is found that, for $\dot M/ \dot M_{\rm Edd} \gtrsim 0.01$ 
(with $\dot M_{\rm Edd}$ = $1.39 \times 10^{18} M/M_{\rm \odot} \rm \ g s^{-1}$),
when the gas flows towards the central  black hole, at a critical radius $r_{\rm d}$ , 
a fraction of the hot gas begins to condense onto the equatorial disc 
plane of the black hole, forming an inner cold accretion disc.
Then the gas will be accreted in the form  of the disc-corona structure extending down to the ISCO 
of the black hole. The size of the inner disc is determined by the initial mass accretion rate.
With the increase of the initial mass accretion rate, the size of the inner disc increases, which results in
both the increase of the Compton reflection scaling factor $\Re$ and the increase of the hard X-ray photon  index $\Gamma$.
By comparing with a sample of Seyfert galaxies with well-fitted X-ray spectra,
it is found that our model can roughly explain the observations.
Our model is briefly described in Section 2. The numerical results and some comparisons with observations 
are shown in Section 3. Some discussions are in Section 4, and the conclusions are in Section 5.

\section{The model}
Initially, the matter is presumed to be accreted in the form of a vertically extended, optically thin,  hot gas (corona)
beyond the Bondi radius of the central black hole. One can see the panel a  of Fig. \ref{fig} for clarity.
We study the interaction between the hot corona and a pre-existing 
geometrically thin, optically thick, cold accretion disc in the vertical direction. 
The interaction between the disc and corona leads to either 
the matter in the accretion disc to be evaporated to the corona or the matter in the corona 
to be condensed to the disc until a dynamic equilibrium is established between the disc and the corona.
We take the vertically stratified method to treat the interaction between the disc and the corona, in which
the accretion flow is divided into three parts in the vertical direction,
i.e., a hot corona, a thin disc,  and a transition layer between them. 
\subsection{The corona}
The corona is treated as a two-temperature hot accretion flow, which is 
described by the self-similar solution of the advection dominated accretion flow (ADAF)
\citep{Narayan1994,Narayan1995}.
The pressure $p$, the  electron number density $n_{\rm e}$, the viscous heating rate $q^{+}$ and the
isothermal sound speed $c_{\rm s}$ as functions of black hole mass $m$, mass accretion rate $\dot m_{\rm c}$,
viscosity parameter $\alpha$ and the magnetic parameter $\beta$
(defined as $p_{\rm m}=B^2/{8\pi}=(1-\beta)p$, where $p=p_{\rm
gas}+p_{\rm m}$, with $p_{\rm gas}$ being the gas pressure and $p_{\rm m}$ being the magnetic pressure)
are expressed as \citep{Narayan1995}, 
\begin{eqnarray}\label{para}
\begin{array}{l}
p=1.71\times10^{16}\alpha^{-1}c_{1}^{-1}c_{3}^{1/2}m^{-1}\dot m_{\rm c} r^{-5/2} \ \ \ \rm g \ cm^{-1} \ s^{-2}, \\
n_e=2.00\times10^{19}\alpha^{-1}c_1^{-1}c_{3}^{-1/2}m^{-1} \dot m_{\rm c} r^{-3/2}\ \ \ \rm cm^{-3},  \\
q^{+}=1.84\times 10^{21}\varepsilon^{'}c_{3}^{1/2}m^{-2}\dot
m_{\rm c} r^{-4}\ \ \rm ergs \ cm^{-3} \ s^{-1}, \\
c_s^2 =4.50\times10^{20}c_{3}r^{-1} \ \ \rm cm^{2} \ s^{-2},
\end{array}
\end{eqnarray}
where $m$ is the black hole mass scaled with the solar mass $M_{\odot}$,  
$\dot m_{\rm c}$ is the coronal/ADAF mass accretion rate scaled with the 
Eddington accretion rate $\dot M_{\rm Edd}$, $r$ is the radius scaled with
the Schwarzschild radius $R_{\rm S}$ (with $R_{\rm S}=2.95\times 10^5 m \ \rm {cm}$), and
\begin{equation}\label{coef}
\begin{array}{l}
{c_1}={(5+2\varepsilon^{'}) \over {3\alpha^2}}g(\alpha,\varepsilon^{'}),\\
\\
{c_3}={2\varepsilon(5+2\varepsilon^{'})\over {9\alpha^2} } g(\alpha,\varepsilon^{'}),\\
\\
{\varepsilon{'}}={\varepsilon\over f}={1\over f} \biggl({{5/3-\gamma}\over {\gamma-1}}\biggr),\\
\\
g(\alpha,\varepsilon^{'})=\biggl[ {1+{18\alpha^2\over (5+2\varepsilon^{'})^{2}}\biggr]^{1/2}-1},\\
\\
\gamma={{32-24\beta-3\beta^2}\over {24-21\beta}},
\end{array}
\end{equation}
with $f$ being the advected fraction of the viscously dissipated energy. 
The energy balance of the electrons in the corona/ADAF is determined by the following equation, 
\begin{eqnarray}\label{energy}
\Delta F_{\rm c}/H=q_{\rm ie} -q_{\rm rad},
\end{eqnarray}
where $\Delta F_{\rm c}$ refers to the flux transfered from the upper boundary of the corona/ADAF to 
the interface of the transition layer. $H$ refers to the height of the corona/ADAF, given by 
$H=(2.5c_{3})^{1/2}rR_{\rm S}$. 
$q_{\rm ie}$ refers to the energy transfer rate from ions to electrons 
\citep{Stepney1983}, which is re-expressed for a two-temperature hot accretion flow as \citep{Liu2002},
\begin{eqnarray}\label{qie}
q_{\rm ie}=(3.59\times 10^{-32} {\rm g\ cm^5\ s^{-3}\ K^{-1}})
n_e n_i T_i {\left(\frac{k T_e}{m_e c^2}\right)}^{-3/2}.
\end{eqnarray}
$q_{\rm rad}(n_{\rm e},T_{\rm e})=$ $q_{\rm brem}+q_{\rm syn}+q_{\rm cmp}+q_{\rm excmp}$ refers to the  
cooling rate of the  electrons in the corona/ADAF, with $q_{\rm brem}$, $q_{\rm syn}$ and
$q_{\rm cmp}$ being the bremsstrahlung cooling rate, synchrotron
cooling rate and the corresponding self-Compton cooling rate
respectively.  $q_{\rm brem}$, $q_{\rm syn}$ and $q_{\rm cmp}$ are
all the functions of  electron number  density $n_{\rm e}$ and
electron temperature $T_{\rm e}$ \citep{Narayan1995}.
$q_{\rm excmp}$ is the Compton cooling rate of the
underling disc photons to the electrons in the  corona/ADAF, which is given by
\begin{eqnarray}\label{cmp}
q_{\rm excmp}={4kT_{\rm e}\over {m_{\rm e}c^2}}n_{\rm
e}\sigma_{T}cu,
\end{eqnarray}
with $u$ being the seed photon energy density of the underlying disc.
$u$ can be expressed as a function of the distance from the black hole, i.e.,
$u={1\over 2}{a T_{\rm eff}^{4}(r)}={1\over 2}{a} \lbrace 2.05T_{\rm
eff,max} ({3 \over r})^{3/4} [1-({3 \over r}
)^{1/2}]^{1/4} \rbrace^{4} $ (with $a$ being the   
radiation constant, $T_{\rm eff, max}$ being the maximum temperature of the accretion disc). 
According to the Spitzer's formula for the thermal conduction \citep{Spitzer1962}, i.e., 
$F_{\rm c}(z)=k_{0}T_{\rm e}^{5/2}dT_{\rm e}/dz$, we simply express the conductive flux transferred from 
the upper boundary of the corona/ADAF to the surface of the transition layer as,
\begin{eqnarray}\label{spizter}
\Delta F_{\rm c} & = & k_{0}T_{\rm em}^{5/2}(T_{\rm em}-T_{\rm
cpl})/H  \nonumber \\
                 & = & k_{0}T_{\rm em}^{7/2}(1- T_{\rm cpl}/T_{\rm
                 em})/H  \nonumber \\
                 & \backsimeq & k_{0}T_{\rm em}^{7/2}/H,
\end{eqnarray}
where $T_{\rm em}$ refers to the electron temperature at the upper boundary of 
the corona/ADAF, which is the maximum temperature of the electrons at a fixed radius. 
$T_{\rm cpl}$ refers to the coupling temperature of the transition layer,  which is much less the 
temperature of the electrons in  the corona/ADAF
\citep{Meyer2007,Liu2007}.
As demonstrated by \citet{Liu2002}, the electron temperature  
in the main body of corona for a fixed radius is approximately a constant, 
we simply replace $T_{\rm em}$ by the local temperature $T_{\rm e}$.

We solve the equations (\ref{para}), (\ref{energy}), (\ref{qie}), (\ref{cmp}) and (\ref{spizter})
for the electron  temperature  $T_{\rm e}$ in the corona by specifying $m$, $\dot m_{\rm c}$, $\alpha$, $\beta$ 
and the maximum effective temperature of the accretion disc $T_{\rm eff, max}$.
Then we can calculate the flux transferred from the corona/ADAF to the transition layer $\Delta F_{\rm c}$ via 
equation (\ref{spizter}). Meanwhile, if it is presumed that the flux at the upper boundary of the corona/ADAF 
is approximately zero, the flux arrived at the interface of the transition layer $F_{\rm c}^{\rm ADAF}$ 
equals to $\Delta F_{\rm c}$.

\subsection{The transition layer}
Following the work of \citet{Liu2007}, the energy balance in the transition layer is determined  by the 
incoming energy transfered from the corona/ADAF, the bremsstrahlung radiation, and the enthalpy carried 
by the mass condensation, which can be expressed as,
\begin{equation}\label{energy-layer}
\frac{d}{dz} \left[\dot m_z \frac{\gamma}{\gamma-1}
\frac{1}{\beta}\frac{\tilde{\Re} T}{\mu} + F_c \right] = -n_e n_i
\Lambda(T),
\end{equation}
with $\tilde{\Re}$ the gas constant.
From equation (\ref{energy-layer}), the condensation/evaporation rate per unit area is given as 
\citep{Meyer2007,Liu2007},
\begin{equation}\label{cnd-general}
\dot m_z= {{\gamma-1} \over \gamma}\beta {{-F_{c}^{\rm ADAF}} \over {\Re
T_{i}/ \mu_{i}}}(1-\sqrt{C}),
\end{equation}
with
\begin{equation}\label{C}
C \equiv\kappa{_0} b \left(\frac{0.25\beta^2 p_0^2}{k^2}\right)
\left(\frac{T_{\rm {cpl}}}{F_c^{\rm{ADAF}}}\right)^2.
\end{equation}
From equation (\ref{C}), setting $C=1$, a critical radius $r_{d}$ is determined. From 
this critical radius inwards, $\dot m_{\rm z}<0$, indicates that the corona/ADAF matter condenses onto
the disc. However,  from this critical radius outwards, $\dot m_{\rm z}>0$, indicates that the matter evaporates
from the disc to the corona/ADAF. 

Currently, we don't consider the potential cooling mechanism such as the Compton cooling  
in the transition layer. Although the Compton cooling in the transition layer should be not 
very important, as analyzed  by \citet{Meyer2007}, it will also increase the radiation of the 
transition layer to some extent.
Consequently, the energy transfered from the corona/ADAF to the transition
layer will be more easily to be radiated out, an increased amount of the  matter will move 
towards the disc plane, leading to a slight increase of the  condensation rate.

\subsection{The disc}
Combining equations (\ref{cnd-general}) and (\ref{C}), the integrated condensation rate in units of
Eddington rate from $R_{\rm d}$ to any radius $R$ of the
disc inward reads,
\begin{equation}\label{condensation}
\dot m_{\rm cnd}(R)= \dot m_{\rm disc}(R) = \int_{R}^{R_d} {4\pi R \over \dot M_{\rm
Edd}} \dot m_z dR.
\end{equation}
According to mass conservation, if the initial mass accretion rate
in the corona/ADAF is $\dot m$,
the mass accretion rate in the corona is a function of distance,
\begin{equation}\label{mdot-corona}
\dot m_{\rm c}(R)=\dot m -\dot m_{\rm cnd}(R).
\end{equation}
The  total  luminosity of the corona/ADAF is derived by integrating the corona/ADAF
region, i.e., 
\begin{equation}\label{Luminosity}
L_{\rm c,t}=\int_{\rm 3 R_{\rm S}}^{R_{\rm d}} q_{\rm rad}
H 4\pi R dR.
\end{equation}

\subsection{The effect of the irradiation of the disc by the corona}
We consider the irradiation of  the accretion disc by the corona/ADAF.
For simplicity, we assume that the radiation of the corona/ADAF can be regarded as a point source 
lying above the  disc center at a height of $H_{\rm s}$. In this case, the reflection scaling factor $\Re$ 
(equivalent to the covering factor) can be defined as follows, 
\begin{eqnarray}\label{R}
\Re&=& {\Omega \over {2\pi}} = {1\over{2\pi}} \int_{R_{\rm in}}^{R_{\rm d}} 
{H_{\rm s}\over {(R^{2}+H_{\rm s}^{2})^{3/2}}} 2\pi RdR \\ 
&=&\bigg[1+\bigg({R_{\rm {in}}\over {H_{\rm s}}}\bigg)^{2}\bigg]^{-1/2}-
\bigg[1+\bigg({R_{\rm {d}}\over {H_{\rm s}}}\bigg)^{2}\bigg]^{-1/2},  \nonumber
\end{eqnarray}
where $R_{\rm {in}}$ and $R_{\rm {d}}$ are the inner boundary and the outer boundary of the accretion disc respectively. 
The effective temperature of the accretion disc can be expressed as follows by including both the 
accretion fed by the condensation and the  irradiation of the corona/ADAF,
\begin{eqnarray}\label{trp}
T_{\rm eff}(r)= 2.05T_{\rm eff,max}^{\prime} \bigg({3 \over r}
\bigg)^{3/4}
\bigg[1-\bigg({3 \over r} \bigg)^{1/2} \bigg]^{1/4} \nonumber \\
\times \bigg[{1+6L_{\rm c,t}(1-a) \over \dot M_{\rm cnd}c^2}
{H_{\rm s} \over 3R_{\rm s}}\bigg]^{1/4} \nonumber \\
=2.05T_{\rm eff,max} \bigg({3 \over r} \bigg)^{3/4} \bigg[1-\bigg({3
\over r} \bigg)^{1/2} \bigg]^{1/4},
\end{eqnarray}
where $a$ is albedo, which is defined as the energy ratio of
reflected radiation from the surface of the accretion disc to incident
radiation upon it from the corona/ADAF, $T_{\rm
eff,max}$ is expressed as ,
\begin{eqnarray}\label{tmaxp}
T_{\rm eff,max}=T_{\rm eff,max}^{\prime} \bigg[{1+6L_{\rm c,t}(1-a)
\over \dot M_{\rm cnd}c^2} {H_{\rm s} \over 3R_{\rm s}}\bigg]^{1/4}.
\end{eqnarray}
$T_{\rm eff,max}^{\prime}$ refers to the maximum effective
temperature from disc accretion which is reached at $r_{\rm
tmax}=(49/12)$. The expression of $T_{\rm eff,max}^{\prime}$ is
given as \citep{Liu2007},
\begin{eqnarray}\label{tmax}
T_{\rm eff, max}^{\prime}=0.2046 \bigg({m\over 10}
\bigg)^{-1/4}\bigg[{\dot m_{\rm cnd(r_{\rm tmax})}\over
0.01}\bigg]^{1/4} \ \rm keV.
\end{eqnarray}

We calculate the condensation rate of the corona/ADAF from equation (\ref{condensation}) and 
the luminosity of the corona/ADAF from equation (\ref{Luminosity})
by specifying the black hole mass $m$, the initial mass accretion rate $\dot m$, 
the viscosity parameter $\alpha$, magnetic parameter $\beta$, 
the albedo $a$, the height of the corona/ADAF $H_{\rm s}$ and a maximum effective
temperature of the accretion disc $T_{\rm eff,max}$. Then a new  
$T_{\rm eff,max}$ is derived by combining the equations (\ref{condensation}), (\ref{Luminosity}), 
(\ref{tmaxp}) and (\ref{tmax}). We make iterations by changing the value of $T_{\rm eff,max}$
until a self-consistent solution of the disc and the corona is found,
including the size of the inner disc $r_{\rm d}$, the condensation rate $\dot m_{\rm cnd}(r)$,
the electron temperature of the electrons in the corona/ADAF $T_{\rm e}(r)$, and the scattering optical
depth of the electrons in the corona/ADAF $\tau_{\rm es}(r)$.
With the derived structure of the disc and the corona in radial direction, we calculated the 
emergent spectrum of the disc-corona system with Monte Carlo simulations, one can refer to  
\citet{Qiao2012,Qiao2013} for details.

\begin{figure*}
\includegraphics[width=95mm,height=37mm,angle=0.0]{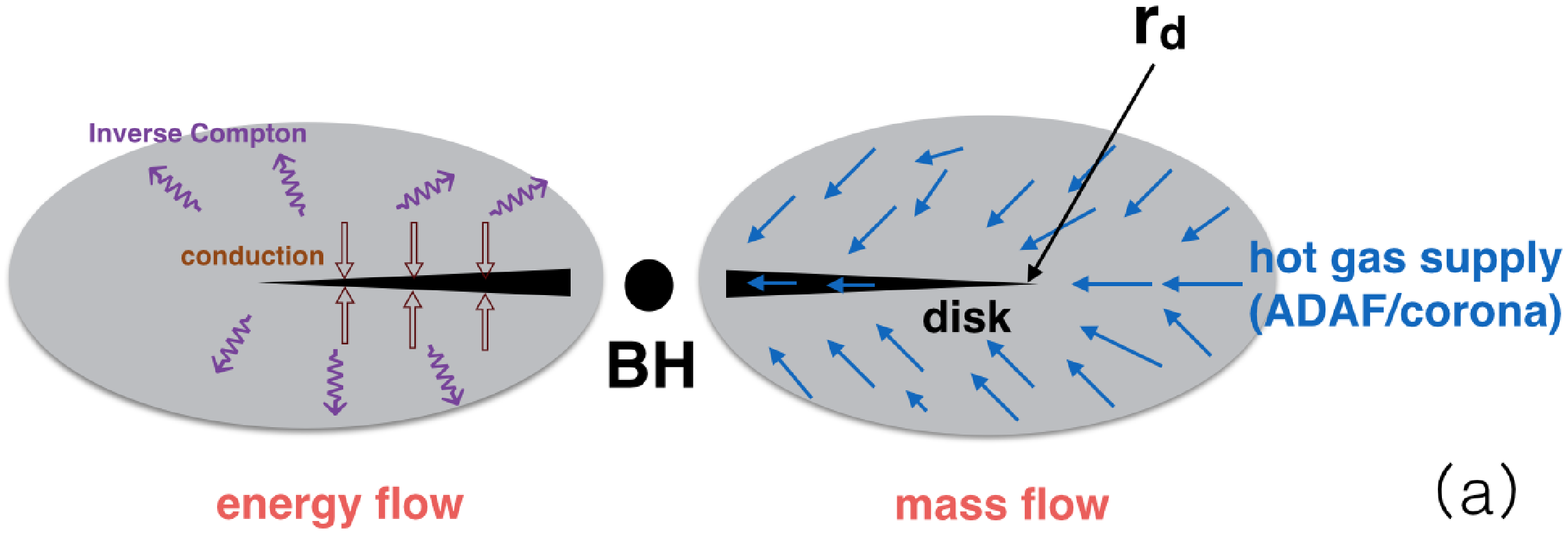}
\includegraphics[width=75mm,height=71mm,angle=0.0]{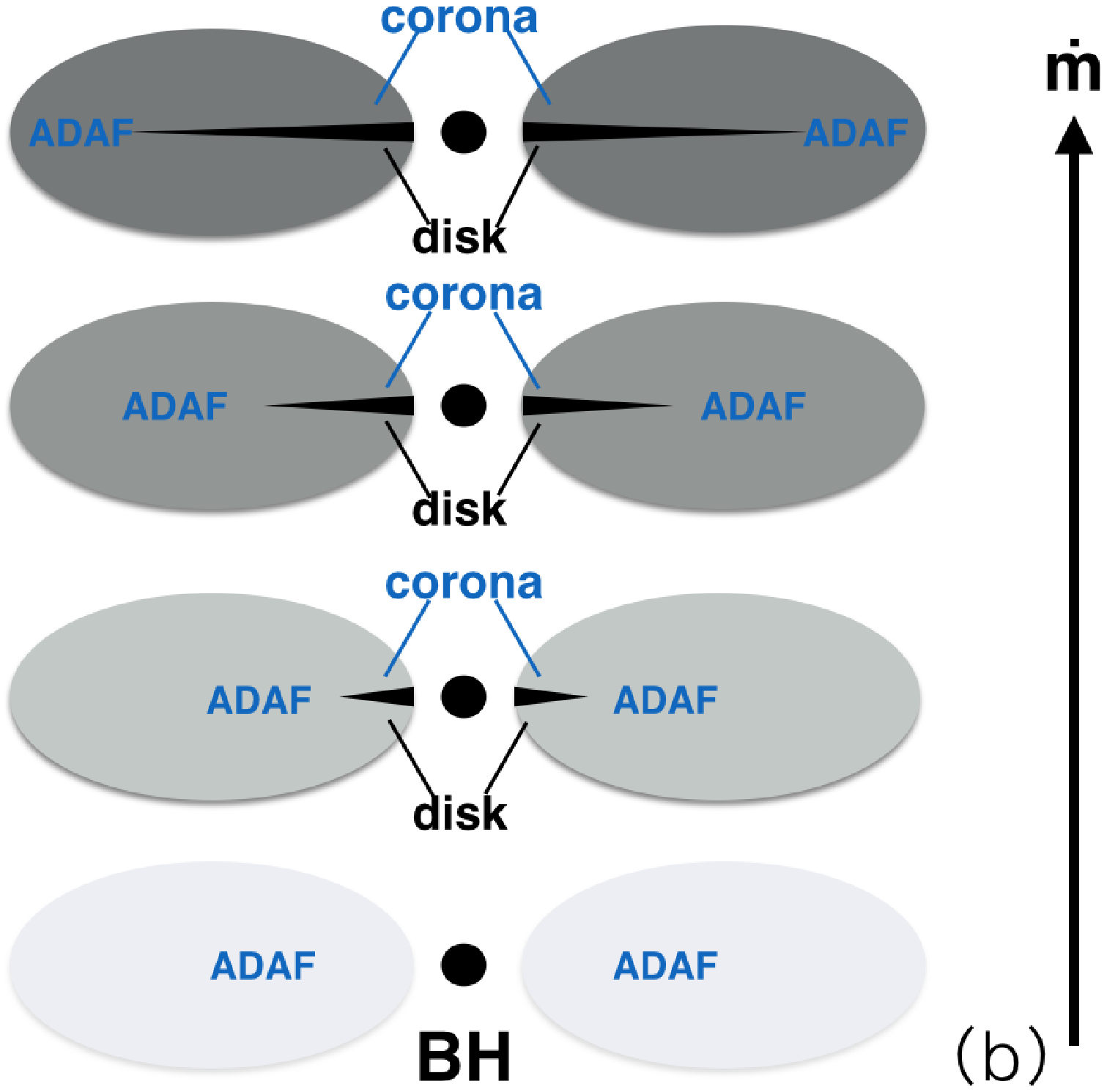}
\caption{\label{fig}}Panel a: Schematic description of the mass and energy flow in the 
disc and corona for the initial mass supply with a vertically extended distribution. 
Panel b: Change of the geometry of the accretion flow with increasing the initial mass
accretion rate from the bottom up.
\end{figure*}

\section{Results}
We solve equations (\ref{para}), (\ref{energy}), (\ref{qie}), (\ref{cmp}),  (\ref{spizter}), (\ref{condensation}) 
(\ref{Luminosity}), (\ref{tmaxp}) and (\ref{tmax}) numerically by specifying the black hole mass $m$,
the initial mass accretion rate $\dot m$, the viscosity parameter $\alpha$, the magnetic parameter
$\beta$, and the reflection albedo $a$.
In this paper, we set $m=10^8$. We set $\alpha=0.3$ as usual 
\citep{King2007}, and $\beta=0.95$ as suggested by MHD numerical simulations \citep{Hawley2001}. 
The reflection albedo is relative low, which is suggested to be $a\sim 0.1-0.2$
\citep[e.g.][]{Magdziarz1995}.
We fix $a=0.15$. The height of the corona/ADAF is difficult to determine, as an example, we fix $H{\rm s}=10R_{\rm S}$.

\begin{figure*}
\includegraphics[width=85mm,height=70mm,angle=0.0]{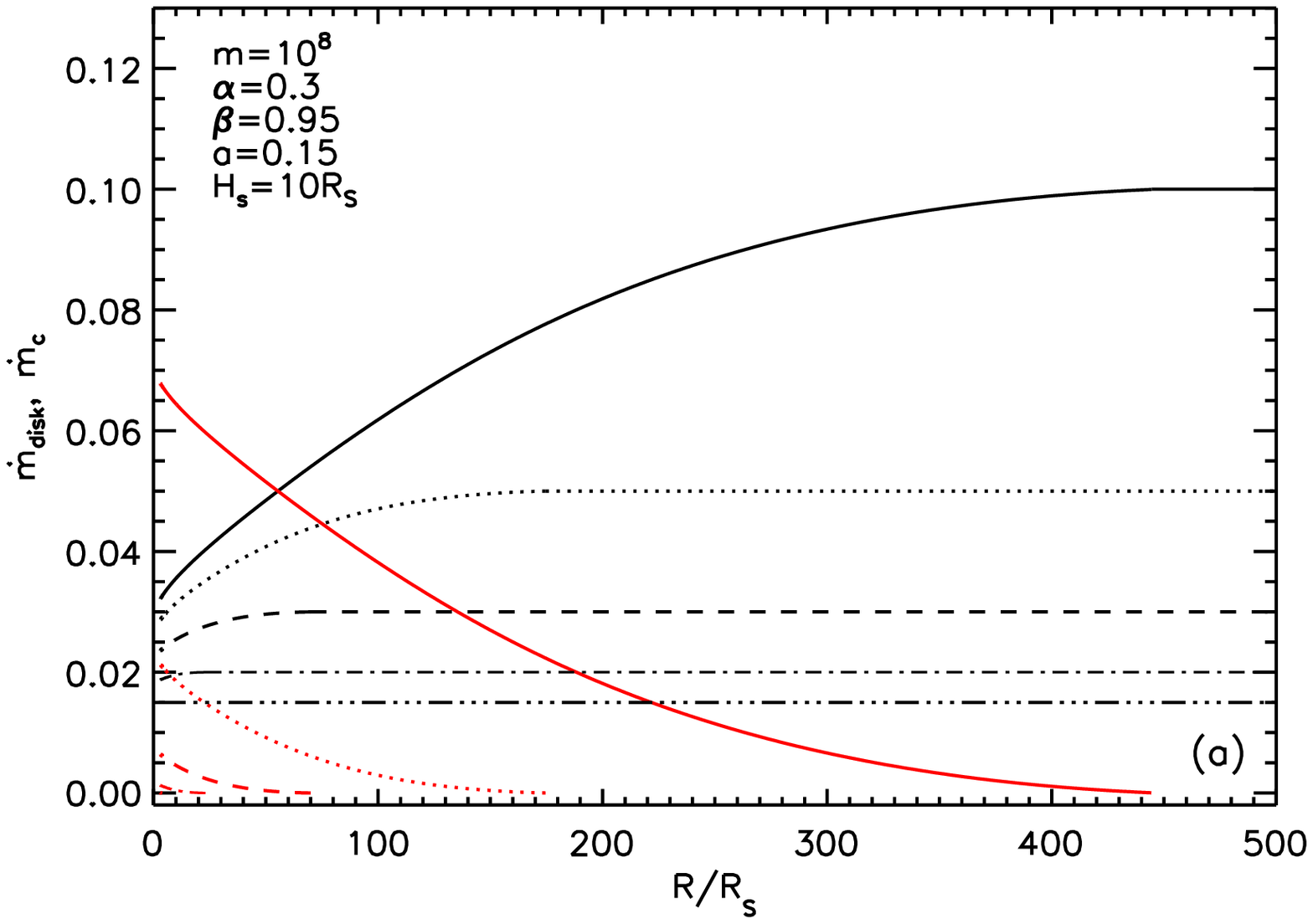}
\includegraphics[width=85mm,height=70mm,angle=0.0]{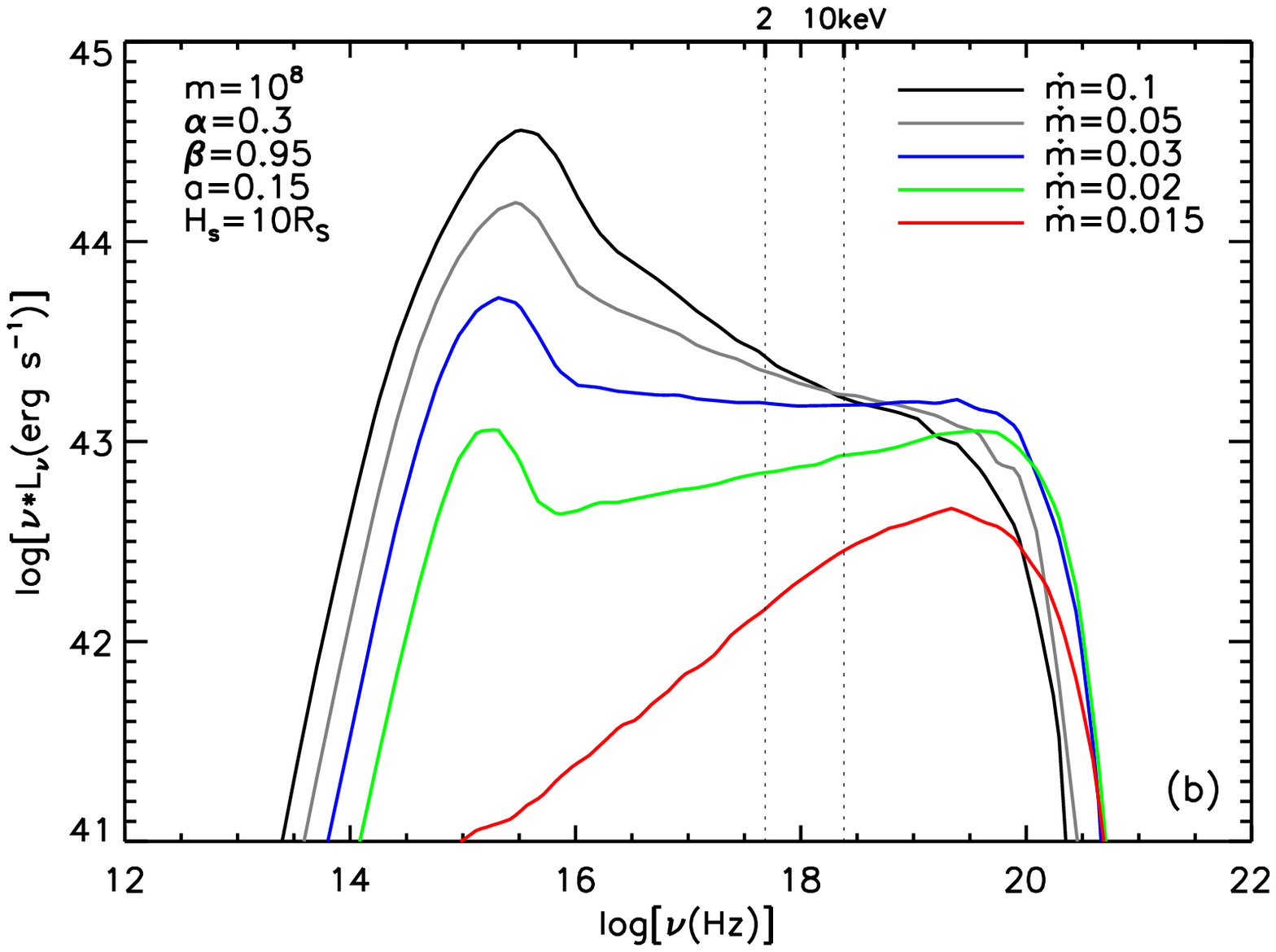}
\caption{\label{mdot}}Panel a: Mass accretion rate in the corona (black line) and 
mass accretion rate in the accretion disc (red line)
as a function of radius. The triple-dotted-dashed line, dotted-dashed line,  
dashed line, dotted line and the solid line are for $\dot m=0.015$, $0.02$, $0.03$,
$0.05$ and $0.1$ respectively. In the calculation, we fix $m=10^8$, $\alpha=0.3$, $\beta=0.95$, $a=0.15$ and
$H_{\rm s}=10R_{\rm S}$ respectively. Panel b: The corresponding emergent spectra. From the bottom up, 
the mass accretion rates are $\dot m=0.015$, $0.02$, $0.03$, $0.05$ and $0.1$ respectively.
\end{figure*}

In the panel a  of Fig. \ref{mdot}, we plot the mass accretion rate in the corona and mass
accretion rate in the disc as a function of radius from the black hole for different initial mass accretion rates.
It is found that the inner disc can be formed when the mass accretion rate $\dot m\gtrsim 0.01$.
For $\dot m=0.015$, the condensation radius is $r_{\rm {d}}=3.75$. With the  increase of the mass accretion rate,
the condensation radius increases, i.e., as examples,
for $\dot m=0.02$, $0.03$, $0.05$ and $0.1$ the condensations radii are $r_{\rm d} = 23$, $70$, $175$ and $444$ respectively.
One can see the panel b of Fig. \ref{fig} for the schematic description for
the change of the geometry of the accretion flow with increasing the initial mass accretion rate.
Assuming that the black hole is a non-rotating Schwarzschild black hole, so the inner boundary of the 
accretion disc $R_{\rm in}$, i.e., the ISCO is $3R_{\rm S}$.
We then calculate the reflection scaling factor $\Re$ with equation (\ref{R}). 
The corresponding reflection factors  are $\Re=0.02, 0.56, 0.82, 0.90$, and $0.94$ for 
$\dot m=0.015, 0.02$, $0.03$, $0.05$ and $0.1$ respectively. 
In the panel b of Fig. \ref{mdot}, we plot the corresponding emergent spectra of the model
(not including the reflection spectra).  From the bottom up,
the mass accretion rates are $\dot m=0.015$, $0.02$, $0.03$, $0.05$ and $0.1$ respectively.  
From the emergent spectra, the hard X-ray photon indices between 2-30 $\rm keV$ are $\Gamma=1.58, 1.88, 2.01, 2.17$, and $2.29$
for $\dot m=0.015, 0.02$, $0.03$, $0.05$ and $0.1$ respectively. 
It is clear that the hard X-ray photon index $\Gamma$ increases with increasing $\dot m$, which is consistent with
the trend observed in luminous AGNs
\citep[e.g.][]{Lu1999,Porquet2004,Wang2004,Shemmer2006,Saez2008,
Sobolewska2009,Cao2009,Veledina2011,Liu2012,Qiao2013,Liu2015}.

\begin{figure*}
\includegraphics[width=85mm,height=70mm,angle=0.0]{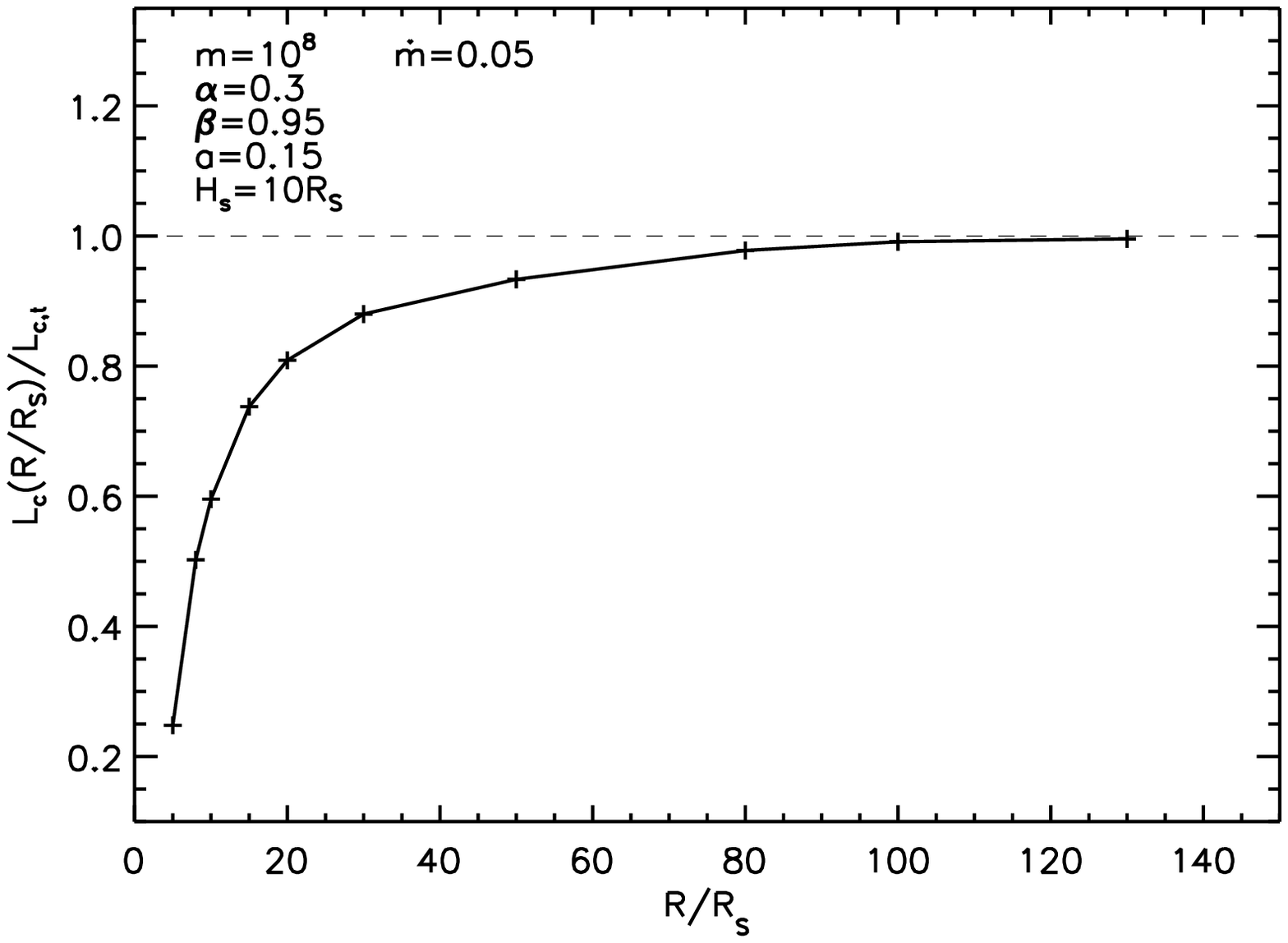}
\caption{\label{compact}}Ratio between the integrated luminosity of the corona/ADAF from 
$R_{\rm {in}}=3R_{\rm S}$ to some radius $R$, $L_{\rm c}(R/R_{\rm S})$,
and the total luminosity of the corona/ADAF $L_{\rm c,t}$ as a  function of radius from the black hole.
\end{figure*}

In our model, in order to simplify the calculation of the irradiation of 
the accretion disc by the corona/ADAF, we  
assume that the corona/ADAF is located as a point source above the disc center at a height of $H_{\rm s}$. 
Actually in our model, initially, the matter is accreted in the form with an extended geometry,
we should check that the luminosity
of the corona/ADAF is indeed compact. In Fig. \ref{compact}, we plot the ratio between the integrated luminosity 
of the corona/ADAF from $R_{\rm {in}}=3R_{S}$ to some radius $R$ and  the total luminosity of the corona/ADAF.
It can be seen that most of the luminosity of the corona is released in a very narrow region. For example, 
$R=10R_{\rm S}$, $L_{\rm c}(R/R_{\rm S})/L_{\rm c,t} \approx 60\%$, 
and for $R=20R_{\rm S}$, $L_{\rm c}(R/R_{\rm S})/L_{\rm c,t} \approx 80\%$,
It is clear that, although the corona/ADAF is geometrically distributed within a wider range in the radial direction, most
of the luminosity of the corona/ADAF is still dissipated in a very compact region, which is roughly consistent 
with our assumption that the corona/ADAF is assumed to be located as a point source above the disc center at the height
of $H_{\rm s}=10R_{\rm S}$.  

\begin{figure*}
\includegraphics[width=85mm,height=70mm,angle=0.0]{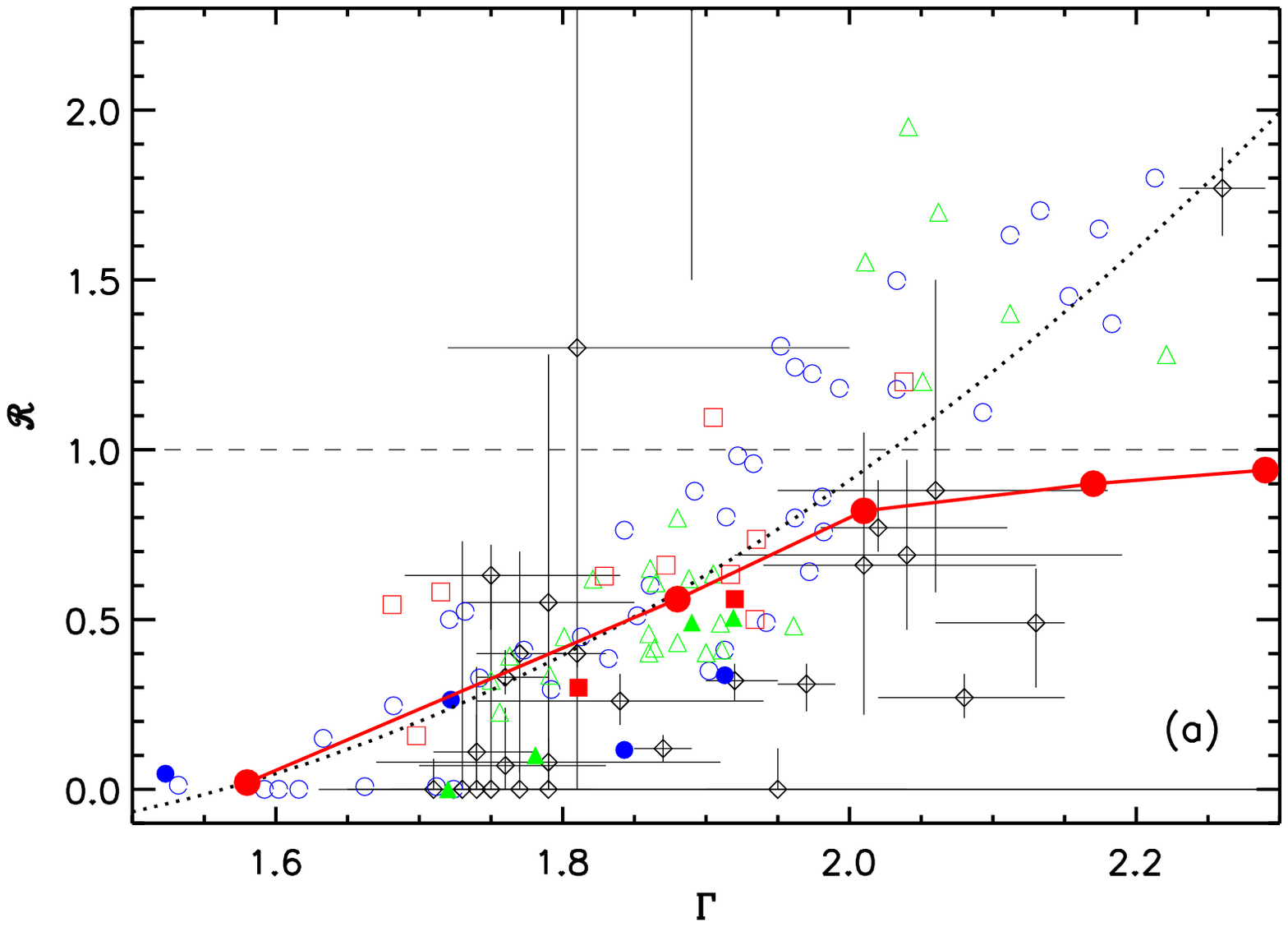}
\includegraphics[width=85mm,height=70mm,angle=0.0]{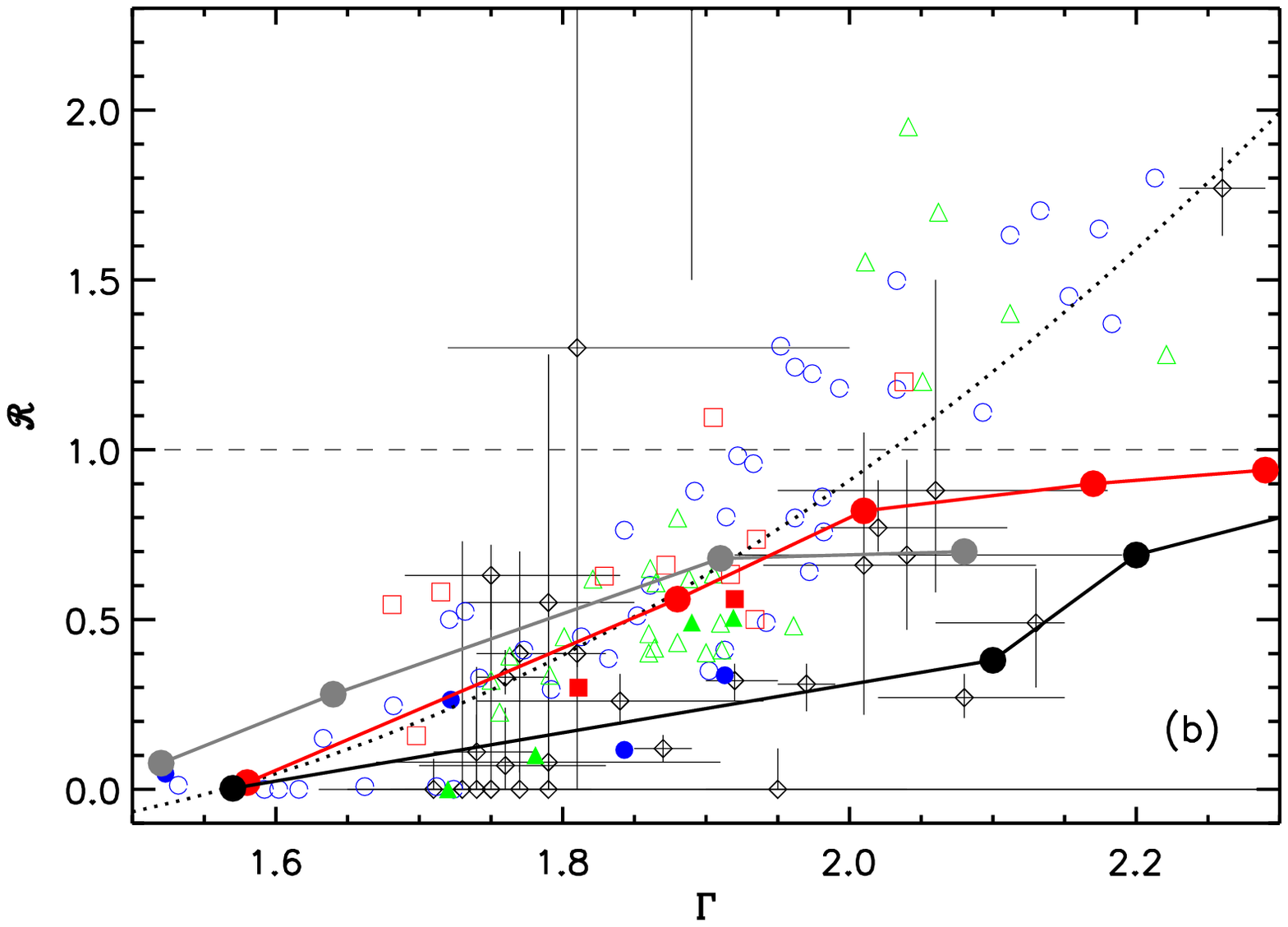}
\caption{\label{gammaR}}Panel a: Correlation between the hard X-ray
photon index $\Gamma$ and the reflection scaling factor $\Re$. 
The black diamond is a sample composed of 28 Seyfert galaxies, including 8 type 1,  
8 intermediate and 12 type 2 Seyfert galaxies, one can refer to \citet{Lubinski2016} for 
details. The colored small points are from a sample composed of 90 Seyfert galaxies  summarized  by 
\citet{Zdziarski2003}, including 11 radio loud AGNs (filled symbols),
and 79 radio quiet AGNs (Open symbols). The blue circles, green triangles and red squares 
correspond to the observations by Ginga, RXTE and BeppoSAX respectively.
The dotted line refers to the best-fitting result with a second-order polynomial.
The red solid line with five  larger red filled circles is our theoretical results, and the 
five larger red filled circles correspond to the model results 
for $\dot m=0.015$, $0.02$, $0.03$, $0.05$ and $0.1$ respectively. In the calculation,
we take $m=10^8$, $\alpha=0.3$, $\beta=0.95$, $a=0.15$ and $H_{\rm s}=10R_{\rm S}$ respectively.
Panel b: All the observational data are same with the panel a. 
We add two sets of theoretical results by assuming different height of the corona.
The solid grey line refers to the case for   
$m=10^8$, $\alpha=0.3$, $\beta=0.95$, $a=0.15$ and $H_{\rm s}=3R_{\rm S}$.
The solid black line refers to the case for   
$m=10^8$, $\alpha=0.3$, $\beta=0.95$, $a=0.15$ and $H_{\rm s}=20R_{\rm S}$.
\end{figure*}

In order to compare with observations for the correlation between $\Gamma$ and $\Re$,
we search for observational data from literatures.
We compiled a sample with 118 Seyfert galaxies by combining two samples from 
\citet{Lubinski2016} and \citet{Zdziarski2003}.
In the sample of \citet{Lubinski2016}, there are 28 Seyfert galaxies, including
8 type 1, 8 intermediate and 12 type 2 Seyfert galaxies.  
In the sample of \citet{Zdziarski2003},  there are 90 Seyfert galaxies, 
including 11 radio loud AGNs, and 79 radio quiet AGNs. 
We fit all the observational data of the 118 sources for the $\Gamma-\Re$ correlation with a 
second-order polynomial. The best fitting result is expressed as follows,
\begin{eqnarray}\label{pfit}
\Re=3.2-5.3\Gamma + 2.1{\Gamma}^2,
\end{eqnarray}
one can see  the dotted line in the panel a  of Fig. \ref{gammaR}.
We plot our theoretical results for the relation between $\Gamma$ and $\Re$ as a comparison. 
One can see the solid red line with five larger red filled circles in the panel a  of Fig. \ref{gammaR}.
The five larger red filled circles from left to right correspond to our theoretical results 
for $\dot m=0.015$, $0.02$, $0.03$, $0.05$ and $0.1$ respectively. 
It can be seen that our model result can roughly match the case for $\Re \lesssim 1$.
We should note that,  since we do not consider the effects such as 
the light bending in the vicinity of the black hole and the relativistic motion of the corona, 
to the strength of the reflection, the current model intrinsically can only be applied to the case for 
$\Re \lesssim 1$.

We test the effect of the height of the corona/ADAF $H_{\rm s}$ to the theoretical relation between $\Gamma$
and $\Re$. One can refer to the  panel  b of Fig. \ref{gammaR}. 
The solid grey line with larger grey circle  points  refers to the case for 
$m=10^8$, $\alpha=0.3$, $\beta=0.95$, $a=0.15$, and $H_{\rm s}=3R_{\rm S}$, 
and the four points from left to right correspond to  $\dot m=0.015, 0.02, 0.05, 0.1$ respectively.
The corresponding condensation radii of the four points from left to right
are $r_{\rm d}=3.7, 6.3, 108, 343$ respectively.
The solid black line with larger black  circle  points  refers to the case for 
$m=10^8$, $\alpha=0.3$, $\beta=0.95$, $a=0.15$, and $H_{\rm s}=20R_{\rm S}$, 
and the three points from left to right correspond to $\dot m=0.014, 0.015, 0.02$ respectively.
The corresponding condensation radii of the three points from left to right 
are $r_{\rm d}=3.4, 26, 63$ respectively.
It is clearly that the assumption for height of the corona/ADAF can affect the 
relation between $\Gamma$ and $\Re$. Theoretically, based on the geometry of the accretion flow in our model, 
we can more accurately determine the relative location between the disc and corona and then the corresponding strength
of the reflection, which is a complicated task to be done in the future,
and beyond the current study. On the other hand, observationally, the height of the corona can be 
roughly estimated by some method,  like the time lag between the emission 
in different X-ray bands, which is also roughly consistent
with the value we adopted in the calculation 
\citep[e.g.][]{McHardy2007,DeMarco2013,Kara2016}.

\section{Discussions}
In the present paper, we proposed a model of the condensation of the corona to explain the observed correlation 
between the hard X-ray photon index $\Gamma$ and the reflection scaling factor $\Re$ in active galactic nuclei,
which is believed to be employed to explore the geometry of accretion flow around a supermassive black hole. 
Historically, several models have been proposed to explain such a correlation.
\citet{Zdziarski1999} proposed a so-called disc-spheroid model to explain the correlation between  
$\Gamma$ and $\Re$. In the disc-spheroid model, an optically thin, hot sphere is surrounded
by a flat optically thick accretion disc, which can extend from a truncation radius,
$d=0$ to $d=\infty$, to the black hole. In this model, the seed photons from the cold 
accretion disc are scattered in the hot sphere, producing the power-law X-ray emission. 
In turn, a fraction of the power-law X-ray emission from the hot sphere is intercepted by the disc.
The intercepted X-ray photons will  be  partly reflected by the accretion disc,
forming the reflection spectrum, and partly be absorbed by the accretion disc, then
reprocessed in the disc as the seed photons to be scattered in the hot sphere.
Finally, a equilibrium feedback between the cold disc and the hot sphere is
established. With the decrease of the truncation radius  of the accretion disc $d$, the solid angle
covered by the accretion disc as seen from the hot sphere increases, leading to a
stronger reflection. Meanwhile, the cooling of the disc to the hot sphere also increases,
leading to a softer X-ray spectrum. By changing $d$ arbitrarily from  $d=0$ to $d=\infty$,
the model can well reproduce the observed $\Gamma-\Re$ correlation for $\Re \lesssim 1$.
We should note that, in \citet{Zdziarski1999}, the author did not consider the self-Compton scattering
of the seed photons (synchrotron radiation and bremsstrahlung) in the hot sphere, 
which may have obvious effects to the X-ray spectra for the truncation radius of the accretion disc
greater than $\sim 30 R_{\rm S}$ \citep[][for review]{Poutanen1997,Esin1997,Poutanen2014}.

Theoretically, the physical mechanism of the truncation of the accretion disc has been studied 
for many years, e.g., the disc evaporation model proposed by 
\citet{Meyer2000b,Meyer2000a}, \citet{Liu2002}, \citet{Qiao2009, Qiao2010}, \citet{Qiaoetal2013},
and \citet{Taam2012}, or the strong ADAF principle by \citet{Narayan1995}, \citet{Abramowicz1995}
and \citet{Mahadevan1997} etc.  
There are two main findings of the previous studies, (1) the truncation radius of the accretion disc 
depends on the initial mass accretion rate $\dot m$, i.e., the truncation radius decreases with increasing $\dot m$;
(2) there exists  a critical mass accretion rate $\dot m_{\rm crit} \sim \alpha^2$.
When $\dot m \gtrsim \dot m_{\rm crit}$, the accretion disc does not truncate, and  extends down to the ISCO
of the black hole. In this case, due to the strong Compton cooling of the seed photons from the 
accretion disc to the corona, the viscous heating of the corona itself can not sustain the existence of the corona. 
The corona collapses very quickly, and the accretion will be dominated by the cold accretion disc.
Consequently the corona  above the disc is too weak to explain the observed power-law hard X-ray emission,
not to mention the reflection, unless another heating mechanism (probably the magnetic reconnection heating) for the 
corona is considered, which is still not very clear \citep[e.g.][]{MeyerHofmeister2012}. 
With these two restrictions, it is unclear whether the correlation between
$\Gamma$ and $\Re$ can still be reproduced well.

Based on the disc evaporation model, \citet{Taam2012} investigated the truncation radius of the accretion disc
as functions of $\alpha$, $\beta$ and $\dot m$. The author  formulized  the expression of the truncation radius as,
\begin{eqnarray}\label{tr}
r_{\rm tr} \approx 17.3 \dot m^{-0.886} \alpha^{0.07} \beta^{4.61}.
\end{eqnarray}
It is clear that the strength of the magnetic field, i.e., $\beta$, is important to the truncation radius of the accretion disc.
Meanwhile, the critical mass accretion rate and the corresponding truncation radius can be expressed as,  
\begin{eqnarray}\label{mcrit}
\dot m_{\rm crit} \approx 0.38\alpha^{2.34}\beta^{-0.41}
\end{eqnarray}
\begin{eqnarray}\label{trmin}
r_{\rm min}\approx 18.80\alpha^{-2.00}\beta^{4.97}.
\end{eqnarray}
By changing $\alpha$, $\beta$ and $\dot m$, especially $\beta$, 
the model can fit the truncation of the accretion disc for most of low-luminosity AGNs and 
the black hole X-ray binaries in there low/hard spectral state  very well 
\citep{Gilfanov2000,Taam2012,Plant2015,DeMarco2015,Basak2016}.
However, we notice that, for example, taking  $\alpha=0.3$ and $\beta=0.95$, the predicted minimum 
truncation radius is  $r_{\rm min}=162$;
taking $\alpha=0.3$ and $\beta=0.8$, the predicted minimum truncation radius is $r_{\rm min}=69$. 
The maximum reflection scaling factors predicted by the above two sets of parameters are 
$\Re_{\rm max}=0.05$ and $\Re_{\rm max}=0.13$ respectively, which are too weak to  explain the strength of 
the reflection in a wider range of $\Re \gtrsim \Re_{\rm max}$.
If an extreme value of $\beta=0.5$ is adopted,  the predicted minimum truncation radius $r_{\rm min}=6.7$, then 
$\Re_{\rm max}=0.8$, which indeed can extend the reflection in a wider range. However,  
as the author stressed, the effect of such a strong magnetic field to the structure of the corona is still unclear,
which probably makes some assumptions in the model not work. For example, in the model 
Spitzer's formula is used for the description of the electron thermal conduction in the corona. 
If a strong magnetic field is considered, Spitzer's formula may not work, which makes the results predicted 
by the model very uncertain \citep{Meyer2002}.  More studies by considering the detailed morphology of 
the magnetic field or other unknown physical mechanism are still needed to reproduce  
smaller truncation radii in the future to extend the strength of the reflection in a wider range.

In this work, throughout the calculation, we assume the central black hole as a non-rotating Schwarzschild 
black hole with the ISCO being $3R_{S}$, and we don't consider the effect of the spin of the 
central black hole to our results.  For an extremely rotating Kerr black hole, the accretion disc can 
extend down to the ISCO being $1/2R_{S}$, which theoretically can increase the strength of the reflection.  
Meanwhile, in the vicinity of the black hole, the effect of light bending can also increase the strength of the 
reflection, which probably can help to explain the case for $\Re \gtrsim 1$ 
\citep[e.g.][]{Fabian2003,Reynolds2003,Dauser2014,Garcia2013}. 
In this work, as the first step, we simply suggested that the condensation of the corona can 
increase the strength of the reflection, then can explain the observed $\Gamma$-$\Re$ 
correlation for $\Re \lesssim 1$. The study  of the effects such as the spin of the black hole, 
the light bending to the reflection is very complicated, which is beyond the scope of the current work.

We should notice that, in the AGNs case, the scattering of the X-ray emission by the outer torus can also
contribute to the reflection. Currently, it is difficult to distinguish the contribution of the disc and torus
to the reflection except for some X-ray very bright sources, such as NGC 4151 \citep{Lubinski2010}. 
Meanwhile, since most of the sources in our sample are type 1 Seyfert galaxies, due to the relative smaller 
viewing angle, the contribution of the torus to the reflection should be not very important.

We suggest that our model can also be applied to the high mass X-ray binaries (HMXBs).  
In HMXBs, the companion star is a massive, bright O or B star.
A fraction of the matter from the  companion star is captured by the compact object via the stellar wind rather
than the Roche lobe overflow (RLO) as in the low mass X-ray binaries (LMXBs) (also called  X-ray transients).
So it is reasonable that, initially, the accreted matter should be in the form of a  vertically, extended 
hot gas in HMXBs rather than be constrained in a very narrow equatorial plane in LMXBs.
Since the hot gas is fully ionized, it can not trigger the ionization instability (viscous instability) 
often observed in LMXBs, which can help us to understand why HMXBs are often relatively stable and the LMXBs are often unstable
\citep[e.g.][]{Lasota2001,Shakura2015}.

\section{Conclusions}
In the work, we proposed a model within the framework of the condensation of the corona/ADAF
to interpret the observed correlation between the hard X-ray photon index $\Gamma$ and the  
reflection scaling factor $\Re$ in AGNs. 
In the model, beyond the Bondi radius the matter is assumed to be accreted in the form 
of a vertically extended, hot gas  by capturing the interstellar medium and stellar wind. 
Within this framework, when $\dot M/ \dot M_{\rm Edd} \gtrsim 0.01$,  at a critical radius $r_{\rm d}$,
a fraction of the accreted  hot gas can condense onto the equatorial disc plane, 
forming a inner disk. Then the accretion will occur in the form a disk-corona system flowing towards the black hole. 
We found that the size of the inner disk increases with increasing the initial mass accretion rate.
With the geometry of the accretion flow, we calculate the reflection scaling factor $\Re$.
Meanwhile, we calculate the emergent spectrum with the structure of the disc and the corona for the 
hard X-ray photon index $\Gamma$.
By comparing with the observations of a sample with 118 Seyfert galaxies, it is found that our theoretical 
relation between $\Gamma$ and $\Re$ can roughly match the observations for $\Re \lesssim 1$. 
Finally, we suggested that the current model can be applied to HMXBs, 
which are believed to be fueled by the hot wind from the bright O or B companion star.


\section*{Acknowledgments}
We thank professor W.M. Yuan from NAOC and Professor R.E. Taam from ASIAA
for very useful discussions and suggestions. 
E.L.Qiao thanks the very useful discussions with Dr. M. Gilfanov 
when visiting Max-Planck Institute for Astrophysics.
This work is supported by the National Natural Science Foundation of
China (Grants 11303046 and 11673026), the gravitational wave pilot B (Grants No. XDB23040100), 
and the National Program on Key Research and Development Project (Grant No. 2016YFA0400804).


\bsp	
\label{lastpage}
\end{document}